\documentclass [preprint,aps,showpacs]{revtex4}
\usepackage[final]{graphics}
\usepackage{amssymb}
\usepackage{amsfonts}
\usepackage{epsfig}
\usepackage{graphicx}

\begin{document}
\title{Absorbing-state phase transitions: exact solutions of small systems}
\author{Ronald Dickman}
\email{dickman@fisica.ufmg.br} \affiliation{Departamento de F\'\i
sica, ICEx, Universidade Federal de Minas Gerais, 30123-970, Belo
Horizonte, Minas Gerais, Brazil}

\begin{abstract}
I derive precise results for absorbing-state phase transitions
using exact (numerically
determined) quasistationary probability distributions for
small systems.  Analysis of the contact process
on rings of 23 or fewer sites yields critical properties (control
parameter, order-parameter ratios, and critical exponents $z$ and
$\beta/\nu_\perp$) with an accuracy of better than 0.1\%; for the exponent
$\nu_\perp$ the accuracy is about 0.5\%.  Good results are also obtained for the
pair contact process.
\end{abstract}

\maketitle

\section{Introduction}

Stochastic processes with an absorbing state arise frequently in
statistical physics \cite{vankampen,gardiner}.  In systems with spatial
structure, phase transitions to an absorbing state, as exemplified by the
contact process \cite{harris,liggett}, are widely studied
in connection with self-organized criticality \cite{socbjp},
the transition to turbulence \cite{bohr}, and
issues of universality in nonequilibrium
critical phenomena \cite{marro,hinrichsen,odor04,lubeck}.
Interest in such transitions should continue to grow in the wake of
experimental confirmation in a liquid crystal system \cite{takeuchi}.
This Letter presents a new theoretical approach to
absorbing-state phase transitions via analysis of exact (numerical)
quasistationary (QS) probability distributions.

The quasistationary probability distribution (QSD)
provides a wealth of information about systems exhibiting an absorbing-state
phase transition \cite{qss,qssim}.
(Since the only true stationary state for a finite system is the
absorbing one, ``stationary-state" simulations
in fact probe QS properties, that is, conditioned on survival.)
In particular, the order parameter and its moments, static correlation functions,
and the QS lifetime are all accessible from the QSD.  Until now, QS properties
of systems with spatial structure have been determined only via simulation
\cite{qssim,qscp,ginelli}; here I develop an effective scheme for
determining the QSD on rings of $L$ sites.

The QSD is defined as follows. Consider a continuous-time Markov process $X_t$
with state $A$ absorbing:
if $X_t = A$, then $X_{t'} = A$ at all subsequent times.
The transition rates
$w_{C',C}$ (from state $C$ to state $C'$) are such that $w_{C,A} = 0$,
$\forall C$.
(Some processes have several absorbing states, $A_1,...,A_n$.)
Let $p_C(t)$ denote the probability of state $C$ at
time $t$, given some initial state $X_0 \neq A$.   The {\it survival
probability} $P_s(t) = \sum_{C \neq A} p_C(t)$ is the probability
that the process has not visited the absorbing state up to
time $t$. We suppose that as $t \to \infty$ the $p_C$, normalized by
the survival probability, attain a time-independent form, allowing us to
define the QSD:

\begin{equation}
\overline{p}_C  = \lim_{t \to \infty} \frac{p_C (t) }{P_s (t)}
,\;\;\;\; (C \neq A),
\label{qshyp}
\end{equation}
with $\overline{p}_A \equiv 0$; it is
normalized so: $\sum_{C \neq A} \overline{p}_C = 1$.

In principle, one could integrate
the master equation numerically and extract the QSD in the long-time limit.
Such an approach is very time-consuming, and essentially useless
for processes with a large state space.  I use instead the iterative
scheme demonstrated in \cite{intme}.  Given some initial guess for the
distribution $\overline{p}_C$, the following relation is iterated until
convergence is achieved:

\begin{equation}
\overline{p}_C' = a \overline{p}_C + (1-a) \frac{r_C}{w_C - r_A} .
\label{iter}
\end{equation}
Here $r_C = \sum_{C'} w_{C,C'} \overline{p}_{C'}$ is the probability flux (in the master equation)
into state $C$ ($r_A$ is the flux to the absorbing state, so that
$1/r_A$ gives the lifetime of the QS state),
and $w_C = \sum_{C'} w_{C',C}$ is the total rate of transitions out of state $C$.
The parameter $a$ can take any value between 0 and 1; in practice
rapid convergence is obtained with $a=0.1$.

The iterative scheme is used to determine the QSD of the {\it contact process} (CP)
on rings of $L$ sites.  In the CP
\cite{harris,liggett,marro}, each site $i$ of a lattice
is either occupied ($\sigma_i (t)= 1$), or vacant ($\sigma_i (t)=
0$).  Transitions from $\sigma_i = 1$ to $\sigma_i = 0$ occur at a
rate of unity, independent of the neighboring sites. The reverse
transition is only possible if at least one neighbor is occupied:
the transition from $\sigma_i = 0$ to $\sigma_i = 1$ occurs at rate
$\lambda r$, where $r$ is the fraction of nearest neighbors of site
$i$ that are occupied; thus the state $\sigma_i = 0$ for all $i$ is
absorbing. ($\lambda $ is a control parameter governing the rate of
spread of activity.)

Although no exact results are available, the CP has been studied
intensively via series expansion and Monte Carlo simulation.
Since its scaling properties have been discussed extensively
\cite{marro,hinrichsen,odor04} we review them only briefly here.  The
best estimate for the critical point in one dimension is $\lambda_c
= 3.297848(20)$, as determined via series analysis \cite{iwanrd93}.
Approaching the critical point, the correlation length $\xi$
and lifetime $\tau$ diverge, following $\xi \propto
|\Delta|^{-\nu_\perp}$ and $\tau \propto |\Delta|^{-\nu_{||}}$,
where $\Delta = (\lambda - \lambda_c)/\lambda_c$ is the relative distance from the
critical point.  The order parameter (the fraction of active sites),
scales as $\rho \propto \Delta^\beta$ for $\Delta > 0$.
Near the critical point, finite-size scaling (FSS) \cite{fisherfss}, implies
that average properties such as $\rho$ depend on $L$ through the scaling
variable $\Delta L^{1/\nu_\perp}$, leading, at the critical point, to
$\tau \propto L^z$, with dynamic exponent $z = \nu_{||}/\nu_\perp$, and
$\rho \propto L^{-\beta/\nu_\perp}$.

The computational algorithm for determining the QSD consists of three
components.  The first (applicable to any model having two states per site)
enumerates all configurations on a ring of $L$ sites.  Configurations differing
only by a lattice translation are treated as equivalent. In subsequent stages
only one representative of each equivalence class is used, yielding a considerable
speedup and reduction in memory requirements.  (The multiplicity, or number of
configurations associated with each equivalence class, is needed for calculating
observables.)  The second component runs through the list of configurations,
enumerating all possible transitions.  Here proper care must be taken to determine
the weight of each transition, due to the varying multiplicity of initial and
final configurations.  The exit rate for each configuration $C$ is simply:
$w_C = n_C + (\lambda/2) c_C$, where $n_C$ is the number of occupied sites and
$c_C$ the number of occupied-vacant
nearest-neighbor pairs.  To determine $r_C$ one enumerates, for each configuration,
all transitions from some other state $C'$ to $C$.  (Each vacant site $i$ in $C$
implies a transition from a configuration $C'$, differing from $C$ only
in that site $i$ is occupied;
each nearest-neighbor pair of occupied sites $i,i\!+\!1$ in $C$ implies
transitions from a $C'$ in which either $i$ or $i+1$ is vacant.  Transitions between
the same pair of configurations $C'$ and $C$ are grouped together, with the proper
multiplicity stored in the associated weight.)  The final part of the algorithm
determines the QSD via the iterative procedure described above.  The specific rules
of the model enter only in the second stage; extension to other models is straightforward.


\begin{figure}[h]
\rotatebox{0}{\epsfig{file=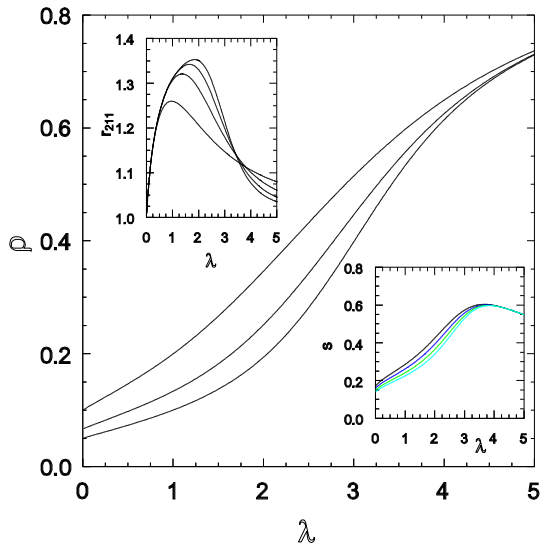,width=10.0cm,height=8.0cm}}
\caption{QS order parameter vs. creation rate $\lambda$ in the CP; system sizes
$L=10$, 15 and 20 (upper to lower). Upper inset: moment ratio $r_{211}$ for system
sizes 5, 10, 15, and 20, in order of increasing maximum value.  Lower inset:
entropy per site $s$ for sizes 17, 19, 21, and 23 (left to right).} \label{qsex1}
\end{figure}

I determined the QSD for the contact process on rings of up to 23 sites.  The number of
configurations scales as $N_c \simeq (2^L -2)/L + 1$ (for $L$ prime this formula is exact).
The number of annihilation transitions is $N_a \simeq 2^{L-1}$ (on average half the sites
are occupied) and that of creation transitions is $N_{cr} \simeq 3 \cdot 2^{L-3}$.
(For $L=23$, there are 364$\,$723 configurations, and $\approx 7.3 \times 10^6$ transitions;
the calculation takes about 16 hours on a 3 GHz processor.)

\begin{figure}
\rotatebox{0}{\epsfig{file=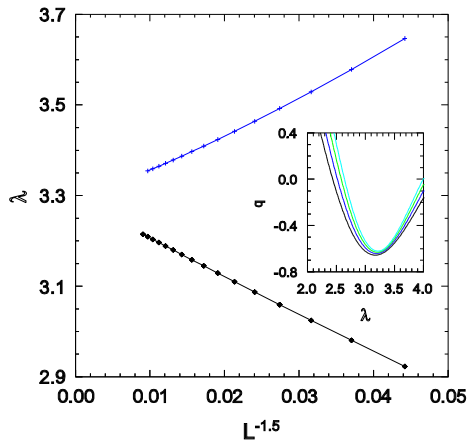,width=9.0cm,height=7.0cm}}
\caption{Contact process: values of $\lambda$ at crossings of $r_{211}$ (upper set), and at
kurtosis minima (lower set) versus $1/L^{1.5}$.
Inset: kurtosis $q$ versus $\lambda$ for (lower to upper) $L=17$, 19, 21, and 23.} \label{cplms}
\end{figure}

The QS order parameter $\rho$
(Fig. \ref{qsex1}), follows the anticipated trend (i.e., $\rho(\lambda)$ is a sigmoidal function),
but does not show any clear sign of the critical point; indeed, no such sign is expected for the small
systems considered here.  A precise estimate
of the critical value $\lambda_c$ can nevertheless be obtained through analysis of
the moment ratios. Let $m_j$ denote the $j$-th moment
of the occupied site density, and $r_{211} \equiv m_{2}/m_{1}^2$.
The values $\lambda_{r,L}$,
marking the crossing of $r_{211}(L)$ and $r_{211}(L+1)$, approach the critical value
systematically, as shown in Fig. \ref{cplms}.  ($\lambda_{r,L}$ is plotted, for
convenience, versus $L^{-1.5}$ as this leads to an approximately linear plot.)
Once preliminary estimates of $\lambda_{r,L}$
(with uncertainty $\sim 10^{-5}$) have been obtained, I perform high-resolution
studies, with $\Delta \lambda = 10^{-4}$, in the vicinity of each crossing; precise
estimates of the crossing values (uncertainty $\sim 10^{-13}$) are then obtained
applying Neville's algorithm \cite{numrec} to the data for $r_{211}(L+1) - r_{211}(L)$.
Using the Bulirsch-Stoer (BST) extrapolation technique \cite{numrec,monroe}, the data
for sizes 8 to 23 furnish $\lambda_c = 3.2961(15)$ and the critical moment ratio
$r_{211,c} = 1.1729(1)$.  These values compare well with the best available estimates
of $\lambda_c = 3.297848(20)$ and $r_{211,c}=1.1736(1)$ \cite{qssim}; the associated
errors are $\approx$ 0.05\%, remarkably small, in light of the system
sizes used.


The moment ratios $r_{3111} \equiv m_3/m_1^3$ also exhibit crossings that converge
to $\lambda_c$.  More surprisingly,
the product $m_{-1} m_1$ exhibits crossings and appears to approach a well defined limit, 1.366(1),
as $L \to \infty$ at the critical point;
simulations ($L=1000$ and 2000), yield 1.374(2) for this quantity.
The reduced fourth cumulant, or kurtosis,
given by $q(\lambda,L)= K_4/K_2^2$, where $K_2 = m_2 - m_1^2$
(the variance of the order parameter) and
$K_4 = m_4 - 4m_3 m_1 -3 m_2^2 + 12m_2m_1^2 -6m_1^4$, does not exhibit crossings but instead
takes a pronounced minimum at a value $\lambda_{qm}(L)$ that converges to the critical value as
$\lambda_c - \lambda_{qm}(L) \propto L^{-1.39(1)}$.  (This property has been verified
in simulations using $L=1000$, and 2000; departures from the minimum
value are evident for $|\Delta| = 5 \times 10{-4}$ \cite{inprog}.)
The sharpness of the minimum, as gauged
by $q'' = d^2 q/d \lambda^2|_{\lambda_{qm}}$, appears to increase rapidly with size: $q'' \propto
L^{1.84(3)}$.  (This is consistent with $q'' \sim L^{2/\nu_\perp}$, as expected from FSS.)
Since a negative kurtosis reflects a probability distribution that is broader at the maximum,
and with shorter tails (compared to a Gaussian distribution with the same mean and variance),
it is natural that $q$ should be minimum at the critical point, where fluctuations are dominant.

The statistical entropy per site, $s = -L^{-1}\sum_j p_j \ln p_j$, is plotted in Fig. 1.
In the large-$L$ limit, $s$ should be zero for $\lambda < \lambda_c$, since the QSD
is concentrated on a set of configurations with vanishing density.  As $L \to \infty$, one
expects $ds/d\lambda|_{\lambda_c}$ to diverge (as is the case for $d\rho /d \lambda$),
and to attain a maximum at some $\lambda > \lambda_c$, approaching zero as $\lambda \to \infty$.
The numerical data are consistent with these trends.


\begin{figure}
\rotatebox{0}{\epsfig{file=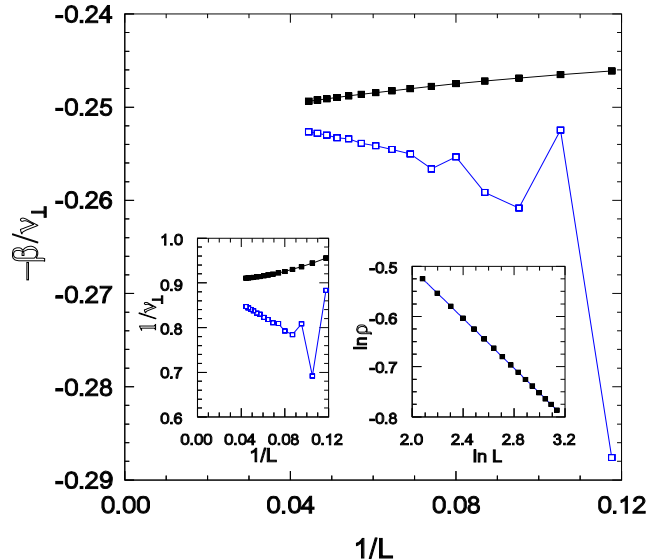,width=12.0cm,height=9.0cm}}
\caption{Main graph: finite differences $\Delta \ln \rho/\Delta \ln L$ versus $1/L$ in the CP (upper set)
and PCP (lower set); left inset: finite differences $\Delta \ln r_1 /\Delta \ln L$ (symbols as in main
graph); right inset: order parameter in CP at $\lambda_c$ versus system size, on log scales.}
\label{rhl}
\end{figure}

Encouraged by the good results of the moment-ratio analysis, I examine three quantities
expected, on the basis of FSS, to exhibit scaling at the critical point:
the QS lifetime $\tau$, the order parameter, and the derivative
$r' \equiv|d r_{211}/d\lambda|$.
(The latter should diverge $\propto L^{1/\nu_\perp}$.)  Despite the small system sizes,
these quantities indeed appear to follow power laws (see Fig. \ref{rhl}).
To obtain precise estimates of the
associated exponents, I calculate the finite difference ratios
$\Delta \ln \rho/\Delta \ln L = [\ln \rho (L) - \rho (L-1)]/[\ln L - \ln (L-1)]$,
(and similarly for $\tau$ and $r'$).  Linear regression of these ratios versus $1/L$,
using the seven largest sizes,
yields $\beta/\nu_\perp = 0.25193(3)$, $z = 1.58054(2)$, and $\nu_\perp = 1.092(1)$.
These estimates differ by 0.06\%, 0.01\%, and 0.5\%, respectively, from the literature values
of 0.25208, 1.5807, and 1.0968 \cite{jensen99}.  Thus data on QS properties of systems with 23 sites or fewer
yield estimates of critical exponents to within half a percent or better!
Precise estimates are also found for $r_{3111}$ and $q$ at the critical point: using BST
extrapolation I find values of 1.5306(5) and -0.5015(5), compared with the simulation
values of 1.526(3) and -0.505(3), respectively \cite{rdjaff}.

To test the robustness of this approach, I apply it to the {\it pair contact process} (PCP).
In the PCP \cite {jensen93},
each site is again either occupied or vacant, but
all transitions involve a pair of particles occupying nearest-neighbor
sites, called a {\it pair} in what follows.
A pair annihilates itself at
rate $p$, and with rate $1-p$ creates a new particle at a randomly
chosen site neighboring the pair, if this site is vacant.
Any configuration lacking a pair of nearest-neighbor occupied sites is absorbing.
Simulation results \cite{jensen93,iwanrdpcp,rdjaff} place the
PCP in the same universality class as the CP (namely,
that of directed percolation).  Unlike the CP, for which quite precise results have been
derived via series expansions, there are no reliable predictions from series or other
analytic methods.

Using, as before, the parameter values associated with crossings of the moment ratio
$m_{211}$, I obtain (for system sizes $L=8$ to 23), the estimate $p_c = 0.07330(3)$,
about 0.3\%  above the best available estimate of 0.077092(1) \cite{pcpdqss}.
The estimates $\beta/\nu_\perp = 0.2483(1)$ and $\nu_\perp = 1.096(2)$, obtained
via the same procedure as used for the CP, are also in good agreement with the accepted values.
Analysis of the QS lifetime however, yields the unexpectedly large
value $z \simeq 2.6$.  In fact, the finite-difference ratios $\Delta \ln \tau/\Delta \ln L$
vary erratically with $L$, indicating that the result for $z$ is unreliable.  This may be
associated with the large number of absorbing configurations in the PCP
(growing exponentially with $L$), so that the extinction rate has not reached its
asymptotic limiting behavior at the system sizes considered here.  Extrapolation of the
moment ratio at $p_c$ yields $m_{122} = 1.174(1)$, and reduced fourth cumulant
$q=-0.500(5)$, again in good accord with the expected values.  (As in the case of the CP,
the value of $\lambda$ at which $q$ takes its minimum approaches $\lambda_c$ with
increasing system size.) Thus the QS properties of the PCP (for $L \leq 23$) permit one to assign
the model to the directed percolation class, despite the lack of a clear result for the
dynamic exponent $z$.

It is natural to inquire whether the QS probability distribution exhibits any simplifying
features.  In an equilibrium lattice gas with interactions that do not extend beyond
nearest neighbors, for example, the probability of a configuration depends only on the
number of particles $N$ and nearest-neighbor pairs $P$.  The the CP,
by contrast, I find that the QS probability of each configuration in a given $(N,P)$
class is distinct (the probabilities typically vary over an order of magnitude or more,
even far from the critical point).  In a broad sense,
this is because, unlike in equilibrium, not all annihilation events possess a complementary
creation event.  For similar reasons, it does not appear likely that the QSD could be obtained
via the maximization of the statistical entropy, subject to some simple set of constraints.

In summary, I show that analysis of exact (numerical) quasistationary properties on relatively
small rings yields remarkably precise results for critical properties at an
absorbing-state phase transition. Deriving the QS distribution involves rather
modest programming and computational effort: the results reported here can be obtained in
a few days on a fast microcomputer.  Applied to the contact process, the analysis yields most
critical properties with an error well below 0.1\%.  For the more complicated PCP, errors are
generally $\leq$ 1\%.  Application to other absorbing-state phase transitions, including some belonging to
other universality classes, is in progress.
The method may also be useful in the study of metastable states, provides a valuable check on
simulations, and may serve as the basis
for phenomenological renormalization group approaches.

\vspace{1em}

\noindent{\bf Acknowledgment}

I thank Robert Ziff for valuable suggestions.
This work was supported by CNPq and FAPEMIG, Brazil.

\bibliographystyle{apsrev}

\begin{thebibliography}{100}
\bibitem{inprog}
        Detailed results on moment ratios will be reported
        elsewhere.
\end{thebibliography}

\end{document}